\documentclass{article}

\usepackage{arxiv}

\usepackage[utf8]{inputenc} 
\usepackage[T1]{fontenc}    
\usepackage{hyperref}       
\usepackage{url}            
\usepackage{booktabs}       
\usepackage{amsfonts}       
\usepackage{nicefrac}       
\usepackage{microtype}      
\usepackage{lipsum}		
\usepackage{graphicx}
\usepackage{doi}

\title{PTHelper: An open source tool to support the Penetration Testing process}

\author{Jacobo Casado de Gracia \\
Universidad Carlos III de Madrid\\
Avda de la Universidad 30\\
Legan\'es, Madrid, Spain\\
\texttt{100488452@alumnos.uc3m.es},
\And 
Alfonso S\'anchez-Maci\'an \\
Universidad Carlos III de Madrid\\
Avda de la Universidad 30\\
Legan\'es, Madrid, Spain\\
\texttt{alfonsan@it.uc3m.es},
}

\renewcommand{\shorttitle}{\textit{arXiv} Template}

\begin{document}
\maketitle

\begin{abstract}
	Offensive security is one of the state of the art measures to protect enterprises and organizations. Penetration testing, broadly called pentesting, is a branch of offensive security designed to find, rate and exploit these vulnerabilities, in order to assess the security posture of an organization. This process is often time-consuming and the quantity of information that pentesters need to manage might also be difficult to handle. This project takes a practical approach to solve the automation of pentesting and proposes a usable tool, called PTHelper. This open-source tool has been designed in a modular way to be easily upgradable by the pentesting community, and uses state of the art tools and artificial intelligence to achieve its objective.
\end{abstract}


\section{Introduction}\label{sec:introduction}
Internet is nowadays one of the core components of society's critical infraestructure. The fast and exponential development of the Internet and hardware and software technologies has also extended the attack surface available to malicious agents. 
Popular examples of this situation are the \textit{WannaCry} \cite{wannacry} or \textit{Mirai} worms \cite{wormsdetect}. Wannacry ransomware affected more than 230,000 devices causing losses over 4 billon dollars . Mirai \cite{wormsdetect} infected more than 300,000 Internet of Thing and embedded devices. Indeed, the attack surface increments in size and depth, as more electronic devices are used, as well as new attack techniques are discovered. 


In order to protect systems 
organizations have developed several security measures to protect their assets and people, including Offensive Security measures such as penetration testing (pentesting). 

Penetration testers are security professionals 
whose objective is to look for flaws in the whole context of an organization, mimicking a real attacker 
in order to fix them and prevent real security incidents. Another of their goals is to evaluate the efficiency of the defense methodologies, policies and tools used by an organization
. In order to achieve their results they use a wide variety of offensive security tools, like network scanners as \textit{nmap} \cite{nmap} or exploit frameworks as \textit{Metasploit}.\cite{metasploit}.
Most of the time, the vulnerabilities found on pentesting assessments are publicly known and stored in databases such as the \textit{National Vulnerability Database} created by MITRE\cite{nvd} or the \textit{Common vulnerabilities and Exposures} created by NIST\cite{cve}.

One of the main problems regarding the pentesting assessments is that the data needed to create and deliver the technical and executive report often lies in distinct and heterogeneous sources and goes through different phases in the pentest. In a brief summary, the data must first be collected during the pentesting assessment, then used in an intelligent way by the pentester to find flaws, to be finally added into the report using a technical and executive format. These actions are very common nowadays and they have transformed the role of the pentester from a technical role into a role that involves organization and knowledge management, apart from technical skills.

Furthermore, pentesting assessments are expensive time and money-consuming activities 
as it is a typically human-driven procedure that requires an in-depth knowledge of the possible attack methodologies and the available hacking tools 
Therefore, automated tools to assist the pentester in his activity are increasingly needed, and often become crucial to the success of the pentest in the short time period previously specified by the scope.


Related works, such as \cite{fidius} and \cite{sarraute}, have demonstrated that some phases of the penetration testing process like the vulnerability Assessment and the Exploitation phase can be automated using techniques such as Artificial Intelligence.

This paper presents a new approach to carry out the automation of the pentests by developing an open-source and modular tool called PTHelper. PTHelper does not try to take over the role of the pentesters as other tools, such as \cite{lore}, does, but to give them support in automating the different phases of the process. It tries to fix the problems presented by the previous approaches, by creating a tool that involves all the phases of the pentesting assessment, including reporting. 


The paper is organised as follows. Section 2 offers an explanation of the context of penetration testing (\textit{pentesting}) and performs an analysis to the state of the art techniques to the problem of automating the process. After that, the approach of PTHelper is presented. In Section 3, the tool's architecture and implementation is detailed 
and in Section 4, experiments are performed in order to check that the objective has been accomplished and to look for points of improvement. Finally, in Section 5, a conclusion and the future work in the development of PTHelper are detailed.

\section{Background}

\subsection{Offensive Security and Penetration Testing}
Nowadays, taking defensive security measures is not enough to ensure the security of modern systems and networks. 
A supplementary proactive strategy, i.e. offensive security, is required where systems are constantly analyzed by professionals that try to imitate a malicious threat actor and detect vulnerabilities before a real threat does. Both offensive and defensive security complement each other to achieve security in an organization. Offensive Security is a broad term that encompasses various techniques and practices, like vulnerability assessments, exploit development, social engineering, and penetration testing, among others.

In penetration testing, vulnerabilities are found and also exploited to compromise other hosts and detect additional vulnerabilities in the infrastructure of an organization. Penetration testers 
act as real adversaries of the organization, so they need to be updated with the latest tools and methodologies in order to mimic the latest attacks.


They have a wealth of resources at their disposal. A primary resource is the National Vulnerability Database (NVD) \cite{nvd} 
that performs an in-depth analysis of software vulnerabilities published in the  Common Vulnerabilities and Exposures (CVE)  database. The  CVE system, an industry standard, assigns a unique identifier to publicly disclosed cybersecurity vulnerabilities, thereby enabling a systematic approach to vulnerability management. The NVD further enriches the CVE data by assigning severity scores to known vulnerabilities using the Common Vulnerability Scoring System (CVSS). 

Beyond just identifying and scoring vulnerabilities, pentesters often need to leverage public exploits to validate these vulnerabilities. Resources like Exploit Database \cite{exploitdb} and Metasploit Framework are commonly used for this purpose. 

These CVE identifiers, CVSS scores, and public exploits play an instrumental role for pentesters. 
The ability to refer to standardized databases and exploit resources simplifies the complex task of vulnerability management, enabling a more precise and targeted approach to maintaining system security.

\subsection{Penetration testing lifecycle}
Penetration testing is a structured process made up of various stages that typically need to be carried out within a limited time (Figure \ref{fig:pentesting_lifecycle}). As mentioned in \cite{eccouncil}, a common methodology to divide the stages is the following:

\begin{figure}[t!]
    \centering
    \includegraphics[width=3.5in]{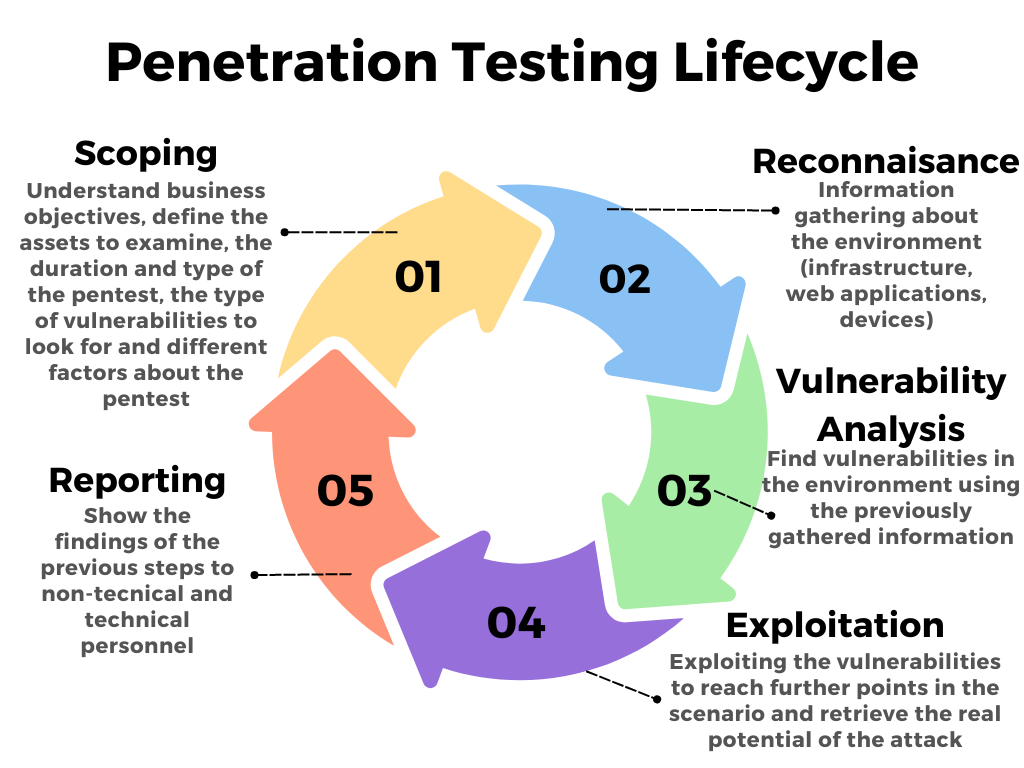}
    \caption{Graphical view of the penetration testing lifecycle.}
    \label{fig:pentesting_lifecycle}
\end{figure}

\begin{enumerate}
    \item Scoping: The pentester works with the client to 
    identify the assets and systems that will be tested 
    and the rules of engagement
    . The purpose of the scope is to ensure that the pentester does not disrupt the expected operations of the client's business.
    \item Reconnaissance: 
    The pentester gathers information about the target environment using specialized tools 
    such as Nmap \cite{nmap}, Nessus \cite{nessus} or by applying special techniques, e.g. Open-Source Intelligence (OSINT). 
    \item Vulnerability Analysis: Vulnerabilities of the network devices found in the last step are identified
    Most of the vulnerabilities are already disclosed publicly and are mapped to vulnerability repositories, like the aforementioned NVD \cite{nvd} vulnerability database. 
    \item Exploitation: In this phase, detected vulnerabilities are exploited 
    . Pentesters can obtain the exploits from several sources 
    including public exploit databases, such as ExploitDB \cite{exploitdb}
    . For \textit{zero-day} vulnerabilities, 
    private exploit databases or manually crafted exploits can be used.
    \item Reporting: This phase 
    consists in the development of an executive report, a high-level overview explaining the findings in detail (the vulnerabilities discovered and their severity), as well as the risks involved and the remediation to avoid them.
    Technical and non-technical sections are included to detail this information to different stakeholders of the organization.
\end{enumerate}

The pentesting lifecycle follows an iterative approach (Figure \ref{fig:pentesting_lifecycle}) as findings and insights from each test may serve as input for subsequent tests, and, sometimes, exploiting a host can lead to space to perform reconnaissance on new hosts that were not previously reachable.

\subsection{Pentest automation}
Automating pentesting can significantly reduce the time and effort to test systems and applications for vulnerabilities. It can also avoid the risk of vulnerabilities or attack vectors being overlooked which may happen in a manual scenario. Automating the report phase provides more time to focus on other phases and helps in keeping track of the information generated throughout the process. 


The automation of penetration testing has been a topic since the last decade. Several studies and approaches have been developed over the course of the years, differing each other in scope and purpose.
The application that first introduced this automation process was made by Boddy et al \cite{boddy}. In 2005, Mark Boddy demonstrated that it was possible to use classical planning to generate hypothetical attack scenarios to exploit a specific system, with the downside of considering only \textit{insider} attacks.

The approach in \cite{sarraute} took into account the presence of \textit{outsiders} or \textit{hackers} and used classical planning to automate the penetration testing process into a framework like Metasploit. One of the goals of this approach was to demonstrate that scalability is not a problem for PDDL solving, as the approach worked great with medium to large networks. Nevertheless, this approach bases its job into having a full overview of the target organization and is focused on the perspective of the defender (not the real-case scenario for a pentester).

The approach of \cite{ghosh} focuses on time efficient generation of a minimal attack graph, using a model-checker that removes visualization problems and avoids state-space explosion. A similar project is \cite{khan}, that tries to automatize the Vulnerability Assessment phase. Both of these projects do not generate the PDDL language automatically and neither of them work with a framework like Metasploit. They present theoretical concepts with complex mathematical calculations.

The FIDIUS framework \cite{fidius} is also one of the first attempts that uses artificial intelligence to automate the penetration testing process. This framework implements DDL for the attack plan generation and also uses a Neural Network to predict the value of a host in terms of the importance for the attacker. The problem is that, as said in \cite{fidius}, "\textit{When using the classical planner the user has to specify all hosts, its connections, its subnets and its services in advance}", which, for sure, does not represent a real pentesting scenario. 

Unlike traditional planning, \cite{durkota} attempts to solve attack graphs with Markov Decision Processes (MDPs), which model the world as states and actions as transitions between states, with a function that assigns a value to each state change. Their work seeks to define an optimal pentesting policy – that is, what best action is – for each state prior to execution based on using a fixed-lookahead horizon. By contrast, the approach in \cite{krautsevich} features an adaptive attacker – i.e., using online techniques alongside an MDP-solving system. Both of these MDP-based approaches assign probabilities to actions, moving uncertainty from the environment into uncertainty in the action’s success. Again, this approach is theoretical and is based on a white-box scenario, which is not the reality when performing pentesting assessments.

These works, despite having some differences in their approach, share several points in common that make them unknown for the pentesting community. The points are the following:

\begin{enumerate}
    \item The scenario in which these approaches work is a white-box scenario, where it is assumed that all the variables of the environment are known by the pentester at first
    . In real scenarios, the pentesters usually need to discover the information by themselves along the assessment, like a real attacker would do.
    \item These studies focus 
    on one specific phase of the pentest, isolating that part from the rest of the 
    asessment. The final result 
    is an automated phase of the pentest, but not a whole automated pentesting process.
    \item These studies 
    develop a mathematical solution, but they do not provide a real and usable tool.
\end{enumerate}

Compared to those, Lore \cite{lore} tried to automate and emulate a red team by using a trained model. However, it has only limited application to the scenarios used for testing as the authors stated: "The systems and software used to train the models are not representative to systems and software in the real world". Additionally, it does not provide support for the reporting phase. 

In contrast to these approaches, this work presents PTHelper, a modular tool designed for pentesters to work in a black-box scenario providing help and tooling for each of the phases of the process. In the next section, the design and architecture of the tool is explained.

\section{Architecture Overview}

\subsection{A different approach to automation}

PTHelper is crafted to provide supportive functionality to the pentester throughout the whole penetration testing assessment. It incorporates four modules, each of them specialized in a different phase of the pentesting process. The novelty of PTHelper lies in the orchestrated interaction of these modules, automating the transition of information from one phase to the next up to the reporting one. 
It is important to note that PTHelper does not render the human expert obsolete. 
It strikes a balance between facilitating a more efficient penetration testing process and maintaining the indispensable human oversight.

\begin{figure}[htb]
    \centering
    \includegraphics[width=3.5in]{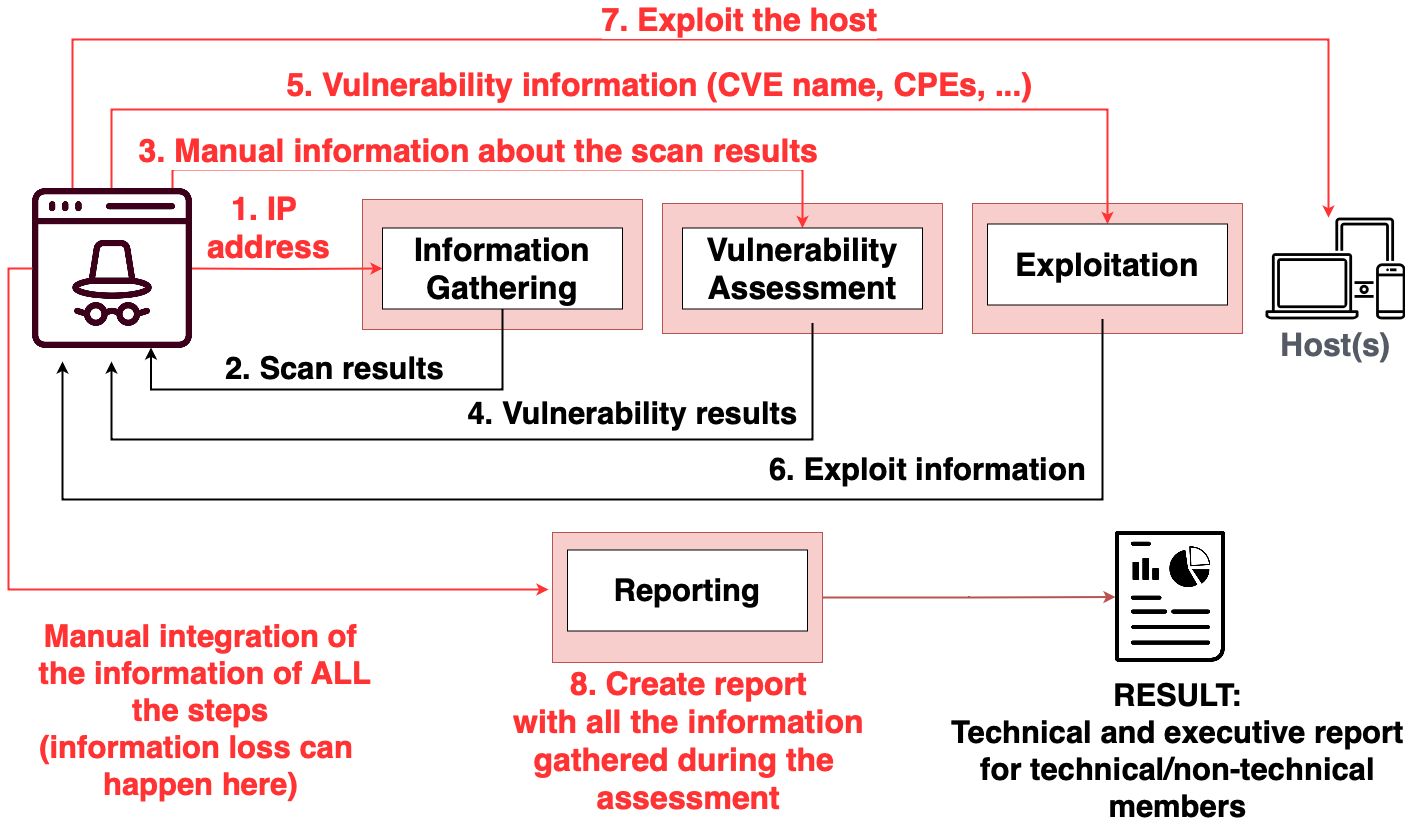}
    \caption{Traditional pentesting methodology. Red lines indicate pentester interactions, black lines indicate the output of each process.}
    \label{fig:traditional_approach}
\end{figure}

\begin{figure}[htb]
    \centering
    \includegraphics[width=3.5in]{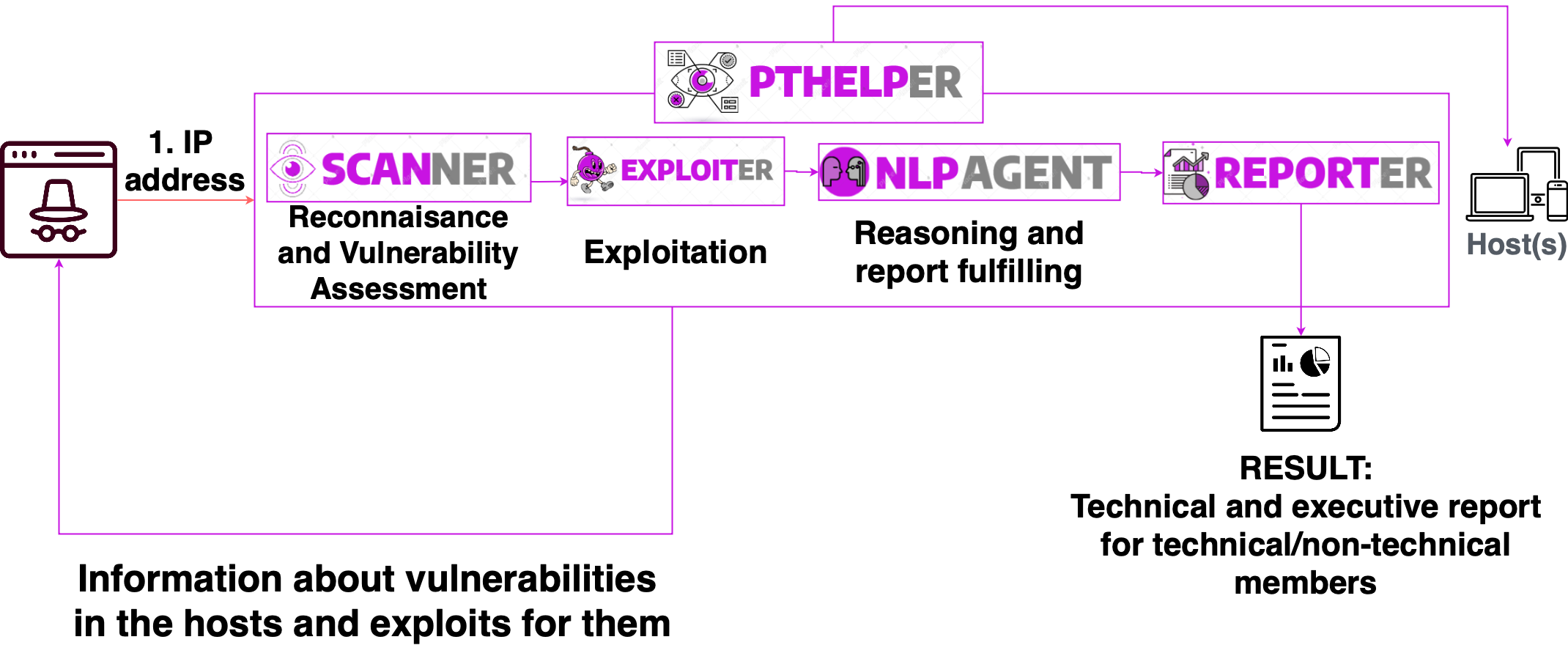}
    \caption{Pentesting methodology proposed by using PTHelper. Red lines indicate pentester interactions, purple lines indicate tool operations.}
    \label{fig:pthelper_approach}
\end{figure}

Figure \ref{fig:traditional_approach} describes the paradigm of the main approaches in the state of the art (note the number of interactions a pentester has to perform) whereas Figure \ref{fig:pthelper_approach} describes the approach of PTHelper regarding the automation of the pentest. Note that both approaches produce the same result, which is an executive report with all the findings obtained during the assessment.

As shown in Figure \ref{fig:pthelper_approach}, PTHelper was designed with the objective of minimizing the pentester's interaction during the phases of the pentest. 
The only provided input during the assessment is the initial information, while the rest of inputs and interactions 
are substituted by PTHelper's native operations and communications between modules, 
reducing the risk of losing information during the different phases.

Initially designed with infrastructure penetration testing in mind, PTHelper possesses the inherent flexibility to accommodate diverse pentesting needs. The architectural design of the tool leverages modularization, which enables it to be responsive to future enhancements and adjustments. Each module can be modified or extended to adapt it to different types of penetration testing, such as Web or Mobile application penetration testing. This adaptive design provides potential to evolve in parallel with emerging pentesting methodologies and challenges in the dynamic field of cybersecurity.




PTHelper leverages specific state of the art tools within each module to fulfill its tasks. 
For instance, the Scanner module currently utilizes \textit{nmap} \cite{nmap} by default. Yet, it allows the integration of other scanners, such as \textit{Shodan}\cite{shodan} or \textit{Nessus}\cite{nessus}, into the Scanner module to meet the unique requirements of different pentesters. 

The forthcoming subsections will offer a detailed exploration of each module incorporated within PTHelper, 
and an explanation of how these modules interact and communicate with each other.

\begin{figure}[t]
    \centering
    \includegraphics[width=3.1in]{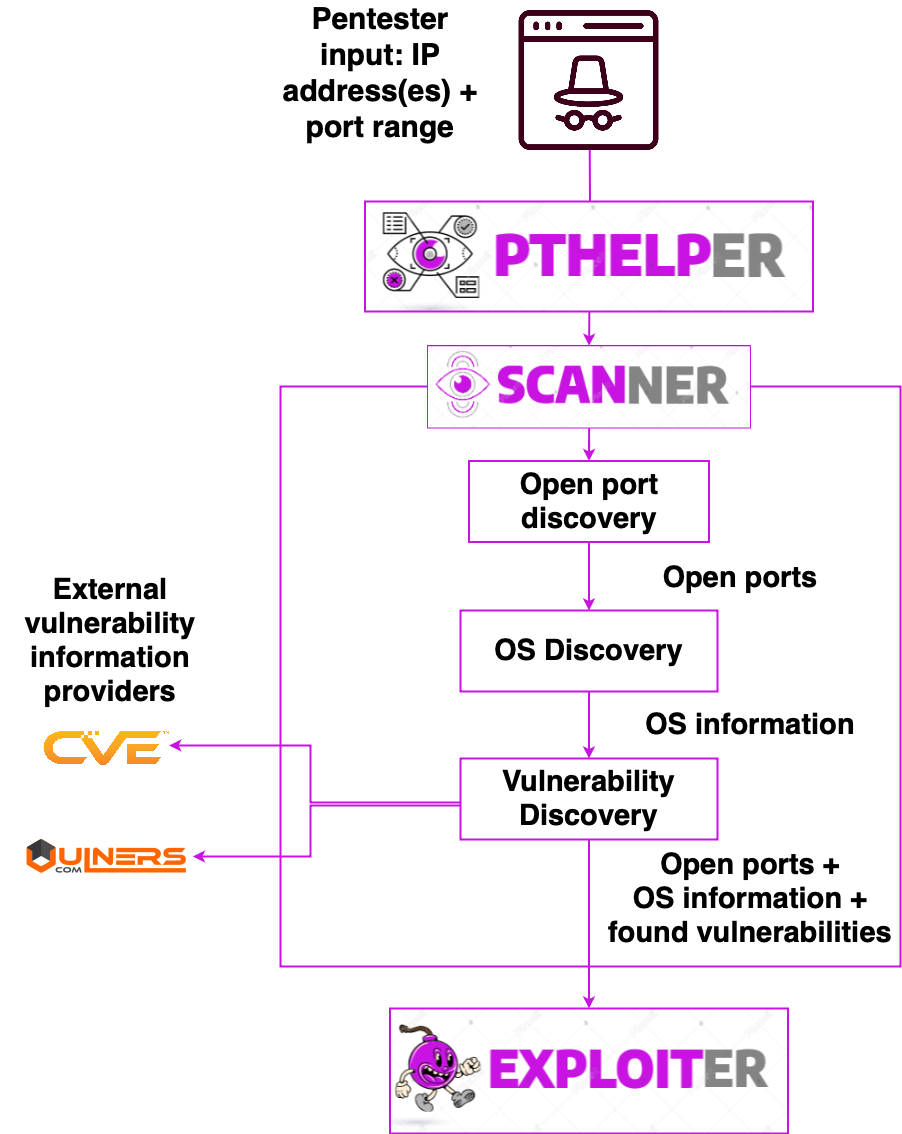}
    \caption{Scanner module operations and interactions}
    \label{fig:scanner}
\end{figure}

\subsection{Scanner module}

The Scanner module is responsible for executing the Reconnaissance and Vulnerability Analysis stages of the pentest. Like most scanning tools, its input is an IP address - the only piece of network infrastructure information required to employ the tool, and usually, the only piece of information available to the pentester. Figure \ref{fig:scanner} provides a visual representation of the mentioned module's operations and interactions.

The primary duty of this module is to extract the maximum possible information about the specified host(s) and relay this information to other modules. 
The module funnels the information to the NLPAgent module. Herein lies a significant difference of PTHelper: instead of the pentester handling the raw scan data, the information is processed and enriched by an AI source.

This module uses the network security scanner \textit{nmap} \cite{nmap} to extract comprehensive data about the specified host(s).
Starting with \textit{nmap}'s Host Discovery scan output, PTHelper identifies open ports on the host and compiles them into a list for further analysis. This list then undergoes into the Service and Version detection scan to obtain detailed insights into the open ports.

For the Vulnerability Analysis phase, the tool utilizes one of \textit{nmap}'s renowned scripts called \textit{vulners}, which is used to extract information about the Common Vulnerabilities and Exposures (CVEs) found in the open ports. Once the information has been enriched with the \textit{vulners} repository, the module uses the \textit{nvdlib} python NVD API wrapper to query the National Vulnerability Database (NVD) \cite{nvd} to 
add the CVE description and CVSS information. 
The \textit{Vulners} database serves as the primary source of CVE information in this script, meaning the most recent CVEs may not be present if the database isn't up-to-date.


\subsection{Exploiter module}

\begin{figure}[t]
    \centering
    \includegraphics[width=3in]{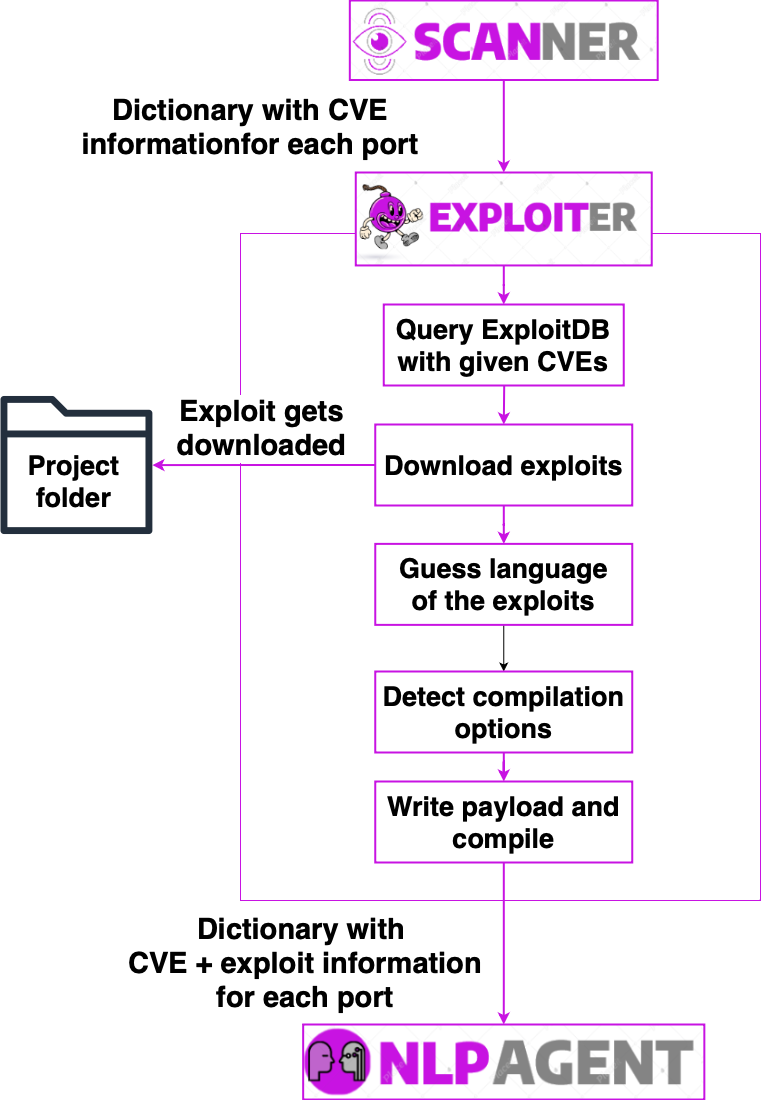}
    \caption{Exploiter module operations and interactions}
    \label{fig:exploiter}
\end{figure}

The Exploiter module 
executes the Exploitation phase of the pentest. 
The list of identified hosts, ports and CVEs, obtained by the Scanner module serves as the foundation for this module's operations. This module uses the given data to search for public exploits corresponding to the identified vulnerabilities of the hosts, subsequently downloading them and providing the pentester with an informative pool of the available exploits. This is the only module that reports the output of its execution to the pentester via the command-line. The displayed output contains a list of the open ports, a brief explanation of each of the CVEs obtained per port as well as the exploits obtained per CVE and the path to the downloaded exploits. 
Figure \ref{fig:exploiter} provides a visual representation of the mentioned module's operations and interactions.

This module interacts with ExploitDB \cite{exploitdb} 
using its API 
to download the required exploits. Additionally, it presents the pentester with essential information, such as the platform of the exploit (e.g., Python, Metasploit, bash) and whether the exploit has been validated by the ExploitDB community, amongst others. This information extends the one provided by the Scanner, therefore obtaining an enhanced overview of the infrastructure. 

The module 
also 
detects the language of the exploit amongst a wide variety of programming languages (whereas the language is written in C, python, Java, Ruby...) by using regular expressions and also detects the compilation options of the script (if they are specified in the exploit) and compiles the script to get a functional exploit that can be used by the pentester.

It's essential to highlight that the Exploiter module doesn't automatically execute the exploits. Instead, this intentional design leaves the final exploit execution in the hands of the pentester. This approach accounts for situations where executing an exploit may inadvertently disrupt the availability of the vulnerable system or where certain modifications within the exploit are required (e.g., altering the IP and port of the target host). Therefore, the pentester retains full control over the final exploit execution, while PTHelper efficiently handles the preparatory steps, significantly simplifying and expediting the exploitation process.

The output of this module includes the information received from the \textit{Scanner} module with the addition of the exploit information. It acts as input for the NLPAgent module and provides an overview of the situation that includes exploit information given the present vulnerabilities.

\subsection{NLPAgent module}


This module integrates a NLP source to fulfill the assessment report generated by the tool with technical and executive insights. The context of the assessment, supplied by the Exploiter module and a set of crafted conversational prompts form the basis for the extraction and generation of this supplemental information. Examples of the generated output vary from the executive report to a severity rationale for each found vulnerability.

In order to fulfill the several parts of the report, a set of prompts that interact with the underlying agent are designed as it will be explained below. 


Regarding the available agents, OpenAI API \cite{chatgpt} is the base NLPAgent that was integrated at this time, due to is ease of use and popularity. Nevertheless, as happens with the rest of the modules, this can be upgradable to include another agents like Google Bard \cite{googlebard} or Llama 2 from Meta \cite{llama2}. The designed prompts that come with PTHelper are reusable between these agents.

In the case of ChatGPT API, the tool is created to be able to change hyperparameters of this agent by modifying a configuration file. This allows the pentester to modify some of the settings of the agent, like the \textit{Temperature} value, which indicates how creative/coherent are the agent's outputs between program executions. Also, the API version of the agent (like ChatGPT API version 3.5 or 4) can be modified in this configuration file. 
Lastly, some techniques that are particular to OpenAI's API were performed in order to reduce token consumption and optimize the token usage without losing information.

\begin{figure}[t]
    \centering
    \includegraphics[width=3in]{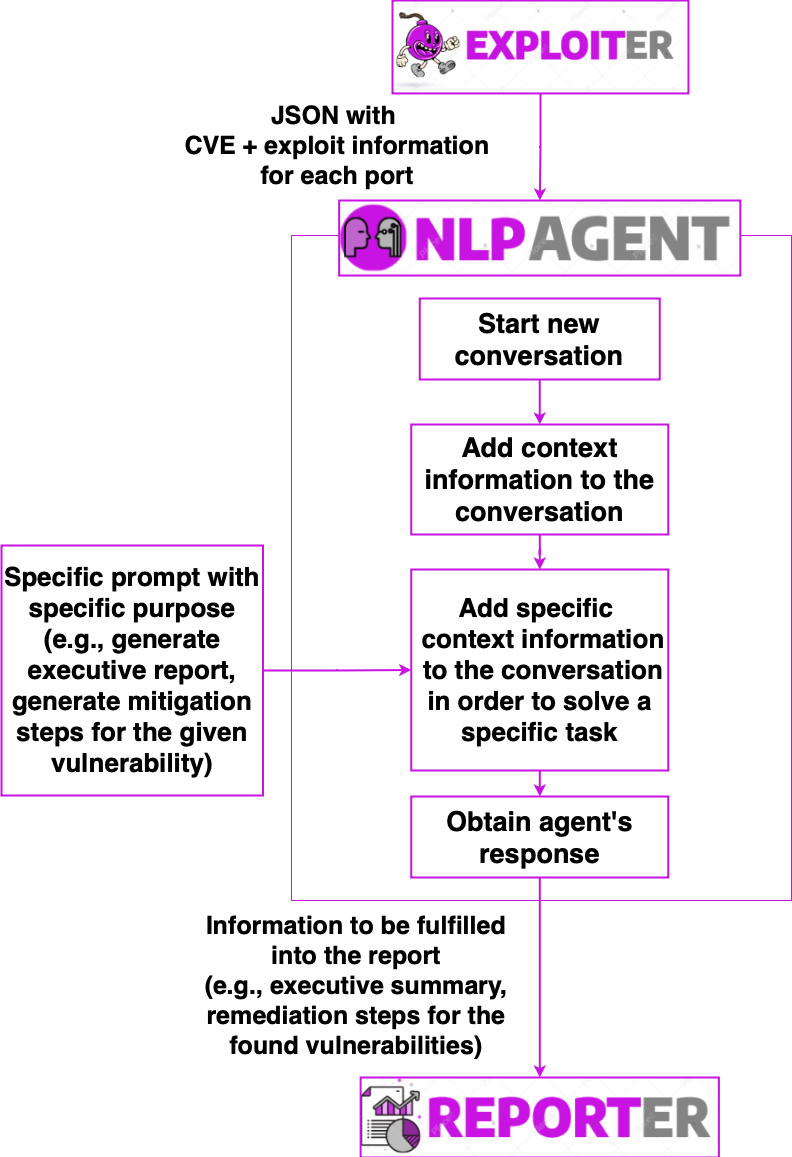}
    \caption{NLPAgent module operations and interactions}
    \label{fig:nlpagent}
\end{figure}

Figure \ref{fig:nlpagent} shows an overview of the module functioning and interactions.

\subsubsection{Applying prompt engineering to the agent}
Each of the parts of the executive report requires a different level of technical content and detail. In order to achieve this, given the same input, which is the one provided by the Exploiter module, adjustments must be made in this module.

A technique called Prompt Engineering is applied. 
Tailored prompts are designed for each of the parts of the report that need to be filled with different expertise levels. A new conversation with the underlying NLP agent is created for each of these report sections, and the tailored prompt is given to the agent as a context message. This means that the agent will not interpret this message as a chat message, but as a message that serves as context and modules its behavior.

Prompts were also used to avoid the generation of misleading content by the underlying agent (such as vulnerabilities that do not exist), which would harm the results and the quality of the whole project. 


To this day, several prompts were crafted in order to fulfill two important parts of the report. The first one is the \textit{Executive Summary} section, that contains information about the hosts and ports that were involved in the assessment, a big overview of the vulnerabilities found in a non-technical language, and a list of mitigations to apply to the vulnerable hosts in order to reduce the risk of the organization. The rest of the prompts are used inside a main function which its main purpose is to generate a \textit{Findings} dictionary. For each of the detected vulnerabilities, additional information is completed using the agent, for example, steps on how to mitigate the vulnerability in the organization to avoid direct attacks, or attach a severity rationale to justify the severity of the given vulnerability. The result is a finding report with valuable information for each of the vulnerabilities found, that completes the information given by the previous \textit{Scanner} and \textit{Exploiter} modules.

The outputs of this module are sent to the \textit{Reporter} module. 


\subsection{Reporter module}

This module executes the Reporting phase of the pentest. It parses the information delivered by the other modules and displays it 
into a SAFR (Security Assessment Findings Report). The SAFR is a complete report that exposes the security findings by the pentester during a certain assessment. 

The module works by using a document template that is filled with the information received from the Reporter and NLPAgent modules. This information includes the executive summary, a table containing an overview of the organization with the scanned IPs, open ports, and found vulnerabilities in CVE format, and a fully-completed \textit{findings} section containing one finding per vulnerability found. The findings contain detailed information about the vulnerability, its severity and a rationale about the impact of the vulnerability in the organization, a list of the public exploits for the given vulnerability and also a remediation list of steps to help with the mitigation of the vulnerability. 

\begin{figure}[t]
    \centering
    \includegraphics[width=3.5in]{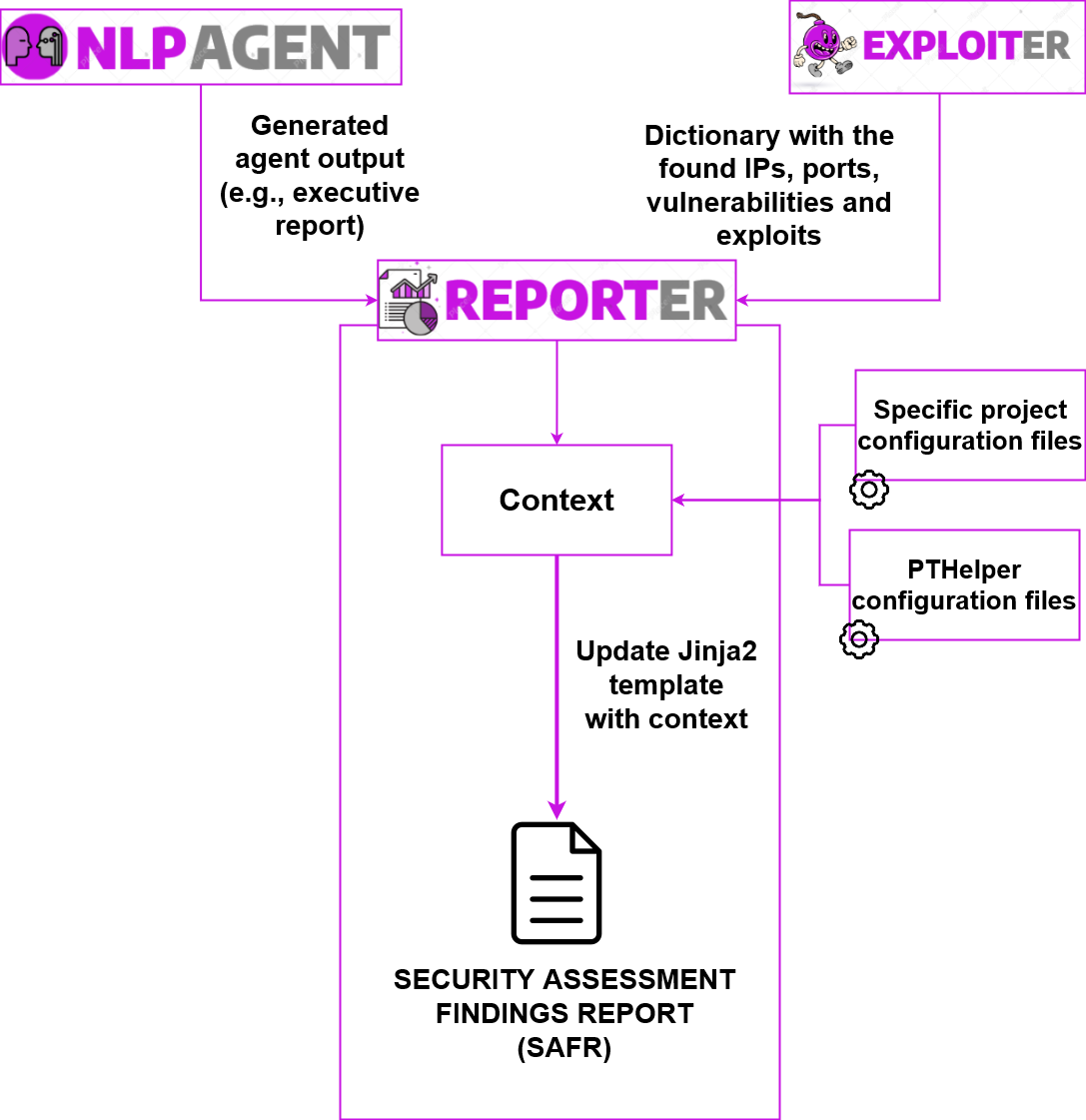}
    \caption{Reporter module operations and interactions}
    \label{fig:reporter}
\end{figure}

Currently the tool includes a base report in \textit{Docx} format. This report comes with all the template tags introduced, ready to be completed with the execution of the tool. The template framework used is the \textit{Jinja2} template framework for \textit{Docx} documents \cite{docxtpl}. This was made with the intention to work with a document that could be easy to edit and visualize. 
Nevertheless, the tool is made to allow pentesters to develop a report in other formats, like Markdown or Latex, and also add additional templating frameworks, to adapt the tool to their needs.

The template that comes with the tool 
includes all the sections that a professional report needs. Some of these sections are ready to be filled by the pentester with additional information to end up with a report that can be used and understood by the technical and non-technical parts of the target organization. Note that the report is not final and it is recommended to review it, and add all the information that is found by manual means.

Figure \ref{fig:reporter} is a visual description of the communications between the Reporter and other modules as described.



\section{Implementation and results}
\subsection{Tool usage and information}
PTHelper is developed in Python3 programming language \cite{python3}, due to its popularity amongst the pentesting communityand it can be downloaded from the GitHub repository\footnote{https://github.com/jacobocasado/PTHelper/}. It requires \textit{nmap} \cite{nmap} binary to be installed in the system (the version of \textit{nmap} which was used during the development and testing process is version 7.94).
In order to install all the Python packages that the tool needs to operate, a \texttt{requirements.txt} file is attached to be fed into the \texttt{pip} command. 

After the requirements are satisfied, PTHelper can be installed as a python package (called \texttt{pthelper}), as configured in the \texttt{setup.py} file. The NVD API key and the OpenAI API key have to be set up in the configuration file (\texttt{config/pthelper\_config.py}) for the modules to work properly. 

PTHelper is usable via the CLI (Command Line Interface), and the following parameters need to be specified:

\begin{itemize}
    \item The IP address (or range of IPs) for the assessment.
    \item The port (or range of ports) to scan of the given IPs.
    \item Type of Scanner that will be used. At this time, only \texttt{nmap} Scanner is available.
    \item Type of Exploiter that will be used. At this time, only \texttt{exploitdb} Exploiter is available.
    \item Type of NLPAgent that will be used. At this time, only \texttt{chatgpt} NLPAgent is available.
    \item Type of Reporter that will be used. At this time, only \texttt{docxtpl} (\textit{Docx} Jinja3 framework) Reporter is available.
    \item Project to store the results. Exploits found by the Exploiter module, the generated report and other results will be placed in the specified project folder. 
\end{itemize}

\subsection{Testing PThelper functionality}
Two simulated pentesting environments were deployed to validate the behaviour of the tool.

\subsubsection{\textit{Black box} infrastructure}
This is a local networking scenario that includes several Virtual Machines simulating a real pentesting environment. Three vulnerable Virtual Machines based on \textit{Metasploitable2} \cite{metasploitable2} and \textit{Metasploitable3} \cite{metasploitable3} were used. These machines contain a wide list of vulnerabilities and are intentionally designed to be vulnerable to help the pentesters develop their skills.

\begin{figure}[t]
    \centering
    \includegraphics[width=3 in]{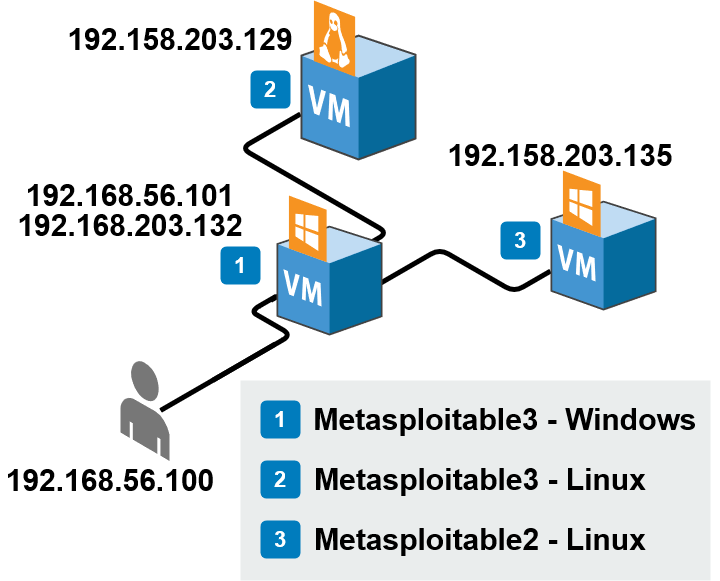}
    \caption{Scenario 1. Black box infrastructure with pivot host.}
    \label{fig:blackbox_infrastructure}
\end{figure}

Figure \ref{fig:blackbox_infrastructure} shows the topology of this environment. The pentester only has network visibility of the first host, a Windows {Metasploitable3} machine (Host A). The other two hosts in the environment are Linux hosts, corresponding to a Linux Metasploitable2 (Host B) and a Linux Metasploitable3 (Host C) machine. The idea is to verify if PThelper can be used in a black-box environment, this is, using PTHelper to compromise a host, and use the tool with the compromised host as pivot with hosts that were not previously accessible. This is a real and typical pentesting scenario.

PThelper was able to detect a RCE (Remote Code Execution) vulnerability on the first host, apart from a wide list of other vulnerabilities. After exploiting the RCE vulnerability (\textit{CVE-2014-3120}) to compromise Host A, a port forwarding technique was used to forward all traffic from PTHelper to the hosts that are not in direct network range (Host B and Host C) to use the tool against these hosts. PTHelper managed to detect a wide list of vulnerabilities in these two hosts and offered a list of exploits, some of which were used to gain DoS (denial of service) and RCE to fully compromise the infrastructure. The output report generated with the tool after performing this experiment by using the tool against the three hosts is also available\footnote{https://bit.ly/pthelper-report}. 

Note that not all the vulnerabilities of these hosts were found in the process, as some of them are web vulnerabilities, that need to be enumerated and exploited using manual means or a specialized web scanner. This was taken into account and it will be one of the main improvement points of the tool.

Table \ref{table:times} details the execution times in seconds of each of the modules of the tool, in order to detect which parts of the tool need more development and optimization for the next versions. The benchmark was executed in a Kali 2023.2 Virtual Machine, with 4GB of RAM memory and 4 CPU cores, and, therefore, the tool could perform better in an environment with more resources.

\begin{table}[h]
    \centering
    \caption{Execution time of PTHelper for each of the hosts (seconds)}
    \begin{tabular}{|l|c|c|c|}
        \hline
        \textbf{Task} & \textbf{Host A} & \textbf{Host B} & \textbf{Host C}\\
        \hline
        Scanner-Port Discovery & 1.16 & 3.10 & 3.13 \\
        \hline
        Scanner-Vulnerability Discovery & 147.15 & 350.28 & 284.39 \\
        \hline
        Scanner - OS Discovery & 0.01 & 0.34 & 0.2 \\
        \hline
        Exploiter & 56.04 & 128.04 & 89.04 \\
        \hline
        NLPAgent - Executive Summary & 158.84 & 289.82 & 233.34 \\
        \hline
        NLPAgent - Finding report & 1454.66 & 2493.24 & 2097.75 \\
        \hline
        Reporter - Render report & 0.1362 & 0.1418 & 0.1591 \\
        \hline
        \textbf{Total time} & \textbf{1720.98} & \textbf{3365.83} & \textbf{2708.75} \\
        \hline
    \end{tabular}
    \label{table:times}
\end{table}

Overall, the most time-consuming part of the Scanner module is the \textit{Vulnerability Discovery} part where queries to the NVD API are performed. For each of these queries, there is a delay when performing and receiving these requests, although the \texttt{delay} parameter was adjusted to the minimum possible allowed by the API. Due to the amount of vulnerabilities per host, this part was time consuming, reaching more than five (5) minutes for the Host B. 

The Exploiter module had an execution time between one (1) and two (2) minutes. 
For the NLPAgent module, times are significantly higher compared to the rest of the modules. Most of the execution time of the tool resides in this module, and, specially, in the Finding Report section of the tool. The justification is simple: The tool performs one query to the OpenAI API per obtained finding. Taking into account that each of the hosts generated more than 20 findings, and that each finding needs to be processed by the engine, the amount of time spent in this operations is high. Finally, the Reporter module does not have a great implication in the execution time as the overall execution time is less than one (1) second.

As the operations of some of the modules depend directly on the found vulnerabilities, the execution time is directly correlated on how vulnerable the host is.

A proper update to the NLPAgent module would be to parallelize the requests performed to the OpenAI API, in order to generate the findings list faster. Also note that the model used was \texttt{gpt-3.5-turbo-16k}. Using \texttt{gpt-4} model will probably provide better values.

\subsubsection{\textit{HackTheBox} machine}

In this experiment, the tool is tested against a host in the Internet instead of the local network. The targeted host is a machine from HackTheBox\cite{hackthebox}. HackTheBox  is a gamified cibersecurity training platform, containing vulnerable machines that can be used to practise hacking skills. A machine of this platform called Blue was used in this experiment.

The tool was used against an instance of this machine, specifying some of the most popular ports as a parameter. The tool managed to discover the vulnerability of this machine, CVE-2017-0144. After discovering the vulnerability, the Exploit module returned several exploits to leverage Remote Code Execution and compromise the host using this vulnerability. One of the obtained exploits was a Metasploit script, which is the one used in the video demonstration\footnote{https://www.youtube.com/watch?v=z7APguceuME} to compromise the host and retrieve the flag, finishing the challenge.

By performing this experiment, it has been possible to demonstrate that the tool can be used by the pentesters in non-local scenarios and that the exploits that the Exploiter module obtains are usable.

\section{Conclusion and future work}
A tool to automate the pentesting process and support the pentested has been presented and tested in two different scenarios. It is reduces the number of the interactions that the pentester has to perform in the assessments in certain types of penetration tests, such as infrastructure penetration testing. The modular design of the tool lets functionality to be expanded to cover other needs of the community. 
This tool is able to cover of the pentesting lifecycle and provide a draft for the integrated report. 

Future work includes fitting in another pentesting scenarios (such as Web or Mobile application pentesting). Another tool improvement, as seen in the experiments section, would be to parallelize some of the operations of the tool, such as the Finding List generation by the \textit{NLPAgent}. The tool has been developed with one option per module, this is, nmap for the Scanner module, ExploitDB for the Exploiter module, OpenAI API for the NLPAgent module and a Jinja2 framework with a \textit{Docx} document for the Reporter module. Additional options will be integrated.


\section{Acknowledgement}
A. Sánchez-Macián would like to acknowledge the support of the R\&D project PID2022-136684OB-C21 (Fun4Date) funded by the Spanish Ministry of Science and Innovation MCIN/AEI/ 10.13039/501100011033.

\bibliographystyle{IEEEtran}

\bibliography{PTHelper}

\end{document}